\documentclass[aps,pre,10pt,
twocolumn,
eqsecnum,
showpacs]{revtex4-1}

\usepackage{feynmp}
\usepackage{amsmath,amssymb}
\usepackage{graphicx}

\usepackage{txfonts}
\usepackage{times}

\newcommand{\esc}{\mathop{\rm esc}\nolimits}
\newcommand{\diver}{\mathop{\rm div}\nolimits}
\newcommand{\rot}{\mathop{\rm rot}\nolimits}
\newcommand{\im}{\mathop{\rm Im}\nolimits}

\def \r {{\bf r}}
\def \s {{\bf s}}
\def \n {{\bf n}}
\def \q {{\bf q}}
\def \ve {\varepsilon}
\def \k {{\bf k}}
\def \bn    {{\bf n}}
\def \bv    {{\bf v}}

\def \be    {{\bf e}}
\def \bs    {{\bf s}}
\def \ki    {\bk^{(i)}}
\def \ks    {\bk^{(s)}}
\def \eperp {\varepsilon_{\perp}}
\def \epar  {\varepsilon_{\parallel}}

\def \H {{\bf H}}
\def \r {{\bf r}}
\def \R {{\bf R}}
\def \s {{\bf s}}
\def \E {{\bf E}}
\def \H {{\bf H}}

\def \n {{\bf n}}
\def \q {{\bf q}}
\def \e {{\bf e}}
\def \ve {\varepsilon}
\def \k {{\bf k}}
\def \K {{\bf K}}

\def \H {{\bf H}}

\def \ki {{\bf k}^{(i)}}
\def \ks {{\bf k}^{(s)}}
\def \bv {{\bf v}}

\begin{document}

\title{Simulation of radiation transfer and coherent backscattering in nematic liquid crystals}

\author{E. V. Aksenova}  \email{aksev@mail.ru}
\affiliation{Saint Petersburg State University, Department of
Physics, Ul'yanovskaya, 1, Petrodvoretz, Saint Petersburg, 198504,
Russia}
 \author{D. I. Kokorin} \email{dmitry@kokorin.org}
 \affiliation{Saint Petersburg State University, Department of
Physics, Ul'yanovskaya, 1, Petrodvoretz, Saint Petersburg, 198504,
Russia}
\author{V.~P.~Romanov} \email{vpromanov@mail.ru}
\affiliation{Saint Petersburg State University, Department of
Physics, Ul'yanovskaya, 1, Petrodvoretz, Saint Petersburg, 198504,
Russia}

\begin{abstract} {
We consider the multiple scattering of light by fluctuations of the director in a nematic liquid crystal. Using  methods of  numerical simulation the peak of the coherent backscattering and the
coefficients of anisotropic diffusion are calculated. The calculations were carried out without simplifying assumptions on the properties of the liquid crystal. The process of multiple scattering was simulated as a random walk of photons in the medium. We investigated in detail the transition to the diffusion regime. The dependence of the diffusion coefficients on the applied magnetic field and the wavelength of light were studied. The results of simulation showed a non-monotonic dependence of the diffusion coefficients on the external magnetic field. A qualitative explanation  of this behavior was suggested using a simple scalar model. For calculation of the peak of the coherent backscattering we used the semianalytical approach as long as in nematic liquid crystals this peak is extremely narrow. The parameters of backscattering peak and of diffusion coefficients which were found in numerical simulations were compared with the experimental data and the results of analytical calculation.
}
\end{abstract}

\date{\today}



\maketitle

\section{Introduction}

The study of multiple scattering of light in liquid crystals  attracts a considerable attention for many
years~\cite{StarkKao1997,Stark1997,vanTiggelenMaynard1996,vanTiggelen1997,HeiderichMaynard1997,
Kuzmin1996,Wiersma2005,Wiersma2004,corsica}.
  The nematic liquid crystals (NLC) have  been studied most thoroughly. Physical properties of these systems are well known and as a rule their elastic and optical parameters are measured with high accuracy.
From the point of view of multiple scattering the liquid crystals are the unique objects. In these systems the multiple scattering occurs on the thermal fluctuations of the orientation. The amplitude and correlation properties of these fluctuations are  studied in detail both experimentally and theoretically. Nematic liquid crystals differ from suspensions where the scattering takes place on the separate particles and also from heterogeneous solid dielectrics where the scattering occurs on the structural inhomogeneities.

The difficulty of studying of multiple scattering in nematic liquid crystals is due to the large optical anisotropy and  the complex structure of the phase function of the single-scattering on the fluctuations of the director.

The most interesting and well-investigated effects of multiple scattering in nematic liquid crystals are the coherent backscattering and the diffusion of light. The effect of coherent backscattering was thoroughly studied theoretically and experimentally for different systems~\cite{vanAlbada1985,Wolf1985,KuzminRomanov1996}, including liquid crystals~\cite{Wiersma2005,Wiersma2004,vanTiggelenMaynard1996,Stark1997,AksKuzRom2009}. Calculation of the backscattering peak is reduced to the summation of ladder and cyclic diagrams. This problem is solved exactly for a system of point-like  scatterers~\cite{mackintosh1988} while for  the scatterers of finite size or fluctuations with finite correlations  length it is necessary to introduce approximations. The  accuracy of these approximations is not always possible to control due to the complexity of the problem being considered.

An effective  description of the radiative transfer in a strongly
inhomogeneous media provides the method of the diffusion approximation. This approach was comprehensively studied in a number of papers~\cite{Steph,Stephen1986,mackintosh1988,Fur,Feng,Zhang,Xzhang,Ger}. The speed of photon diffusion and diffusion coefficients are obtained in the experiments on  the passage of  the short pulses through the medium~\cite{Genack}.
In isotropic systems the diffusion coefficient is determined by the relation
$D=vl_{tr}/3$, where $v$ is the velocity of light in the inhomogeneous medium and $l_{tr}$
is the transport length. In absorbing media  the ratio
$D={v}/[3(\mu'_s+\mu_a)]$ is used for the diffusion coefficient $D$ where~$\mu'_s$ and $\mu_a$ are the coefficients of
scattering and absorption. The correctness of this expression was discussed in the series of papers~\cite{Chance,Dur,Aron}.

Considerable interest has been attracted to the study of the diffusion of photons in
nematic liquid crystals. This problem has been discussed in~\cite{vanTiggelenMaynard1996,Stark1997,mackintosh1988,Wiersma2004,Wiersma2005,Copic}. In particular in the
works~\cite{Stark1997,StarkKao1997}, the analytical
method for calculation of the diffusion tensor has been proposed. The components of the diffusion tensor were measured in
~\cite{Wiersma2004, Wiersma2005, StarkKao1997} by the pulse method  and by the deformation of the light beam
passing through the sample. However    the detailed studies of the dependence of the diffusion coefficients
on external fields and parameters of the system due to the complexity of the
experiments were not carried out. It creates  difficulty in the detailed comparison of the  theoretical and  experimental results.

On the other hand at the present time there exist
numerical simulations methods that allow to
overcome many difficulties occurring in the analytical approach.
In fact this numerical method can be considered as simulation of
the real experiment. The advantage of this approach is that in this case it is possible to extract detail information on the process under consideration and in particular to establish
the accuracy of various approximations introduced in analytical
studies.

In this paper we study the diffusion of photons in
oriented nematic liquid crystal. The transition from multiple scattering to the diffusive regime was analyzed in detail. The dependence of the diffusion coefficients on the orienting magnetic field and on the wavelength of light was investigated. Non-monotonic dependence of the diffusion coefficients on the external magnetic field has been found and qualitative explanation of this effect is suggested.

Besides we study the light diffusion the coherent backscattering in NLC  by numerical simulation. Similar calculations were carried out in a number of papers~\cite{HeiderichMaynard1997,Wiersma2005,KuzminValkov2011}. In these articles the peak of the coherent backscattering was obtained and its behavior under various conditions were analyzed. However due to the complexity of the problem various simplifying assumptions were introduced such as the one-constant approximation, the independence of the extinction coefficient of the direction of the beam, and etc. In our work we make a simulation based on actual parameters of the liquid crystal without  introduction of any simplifying assumptions. It allowed us to perform a quantitative comparison of the numerical results with the
experiment~\cite{Wiersma2005,Wiersma2004} and with analytical calculations~\cite{AksKuzRom2009}.

The paper is organized as follows. The first and second sections summarize the basic equations
describing the fluctuations in a nematic liquid
crystal, its optical properties and single and multiple scattering. The third part describes  the method of numerical simulation.
The fourth part contains  description of the transition to the diffusion approximation and discusses of the dependence of the diffusion tensor on the magnetic field. The fifth part contains the results of numerical simulation of coherent backscattering. The main results of the work are presented in the conclusion.

\section{Propagation of light in an anisotropic fluctuating medium}

We  consider  NLC in an external magnetic field . The orientation of the liquid crystal is determined by the unit vector director $\n(\r)$. The free energy of distortion has the form~\cite{Dj}
\begin{multline}
 \label{1.1}
 F=\frac 12 \int d\r \{K_{11}(\diver\n)^2+
 K_{22}(\n\cdot \rot\n)^2
 \\ +K_{33}(\n\times\rot\n)^2
 -\chi_a (\n\cdot\H)^2\}.
\end{multline}
Here $K_{ll}$, $l=1,2,3$ are Frank modules,
$\chi_a=\chi_\parallel-\chi_\perp$, $\chi_\parallel$, $\chi_\perp$ are the magnetic susceptibilities along and normal to $\H$, $\H$ is the
constant external magnetic field. It is assumed that the sample is large enough so that we can ignore the interaction energy with the anchoring surface. We consider that
$\chi_a>0$. In equilibrium the vector director $\n=\n^0$ is constant and
for $\chi_a>0$ it is directed along the magnetic field, $\n^0\parallel \H$.

As an optical system the nematic liquid crystal is  uniaxial
with a permittivity tensor
\begin{equation}
 \label{2.1}
\ve_{\alpha\beta}(\r)=\ve_\perp\delta_{\alpha\beta}+
\ve_an_\alpha(\r)n_\beta(\r),
\end{equation}
where $\ve_a=\ve_\parallel-\ve_\perp$, $\ve_\parallel$, $\ve_\perp$ are
permittivities  along and across the director $\n^0$.

Eigenwaves $\E^0(\r)$ in a uniaxial media are two plane
waves, i. e., ordinary, (o), and extraordinary, (e), with wave vectors
$\k^{(o)}$ and $\k^{(e)}$
$$
\E_{(j)}^0=E_{(j)}^0\e^{(j)}e^{i\k^{(j)}\cdot\r}, \quad j=o,e.
$$
Here $E_{(j)}^0$ is the amplitude of the field, $\e^{(j)}$ is the unit
polarization vector, $k^{(j)}=k_0n^{(j)}$,
\begin{equation}
 \label{2.6}
 n^{(o)}=\sqrt{\ve_\perp},\quad n^{(e)}=
 \sqrt{\frac{\ve_\parallel\ve_\perp}{\ve_\perp+\ve_a\cos^2\theta_e}}
\end{equation}
are refractive indices of the ordinary and the extraordinary waves, $\theta_j$ is the angle between the vectors $\n^0$ and $\k^{(j)}$.

When describing the  wave propagation in an inhomogeneous medium it is convenient to present the wave
equation in the integral form
\begin{equation}
 \label{2.8}
 \E(\r)=\E^0(\r)+ k_0^2\int d\r_1\hat G^0(\r,\r_1)
 \delta \hat\ve(\r_1)\E(\r_1),
\end{equation}
where $\delta\hat\ve(\r)$ are the fluctuations of the permittivity tensor,
$\hat G^0(\r,\r_1)$ is the Green's function of the electromagnetic field.

In the far zone  the Green's function of the uniaxial medium
has the form in the coordinate representation~\cite{LN73,LN76}
\begin{equation}
 \label{2.11}
 \hat G^0(\R) =\frac{k_0^2}{4\pi R}
 \sum\limits_{j=o,e}n^{(j)}f_{(j)}\frac{\e^{(j)}\otimes \e^{(j)}}
  {(\e^{(j)}\hat\ve^0\e^{(j)})}\exp(i\k_{st}^{(j)}\cdot\R),
\end{equation}
Here
\begin{align}
 \label{2.12}
 \k_{st}^{(o)}&=\sqrt{\ve_\perp}k_0\frac{\R}{R}, \\
 \k_{st}^{(e)}&=
 k_0\left[\frac{\ve_\parallel\ve_\perp}{(\R(\hat\ve^0)^{-1}\R)}
 \right]^{1/2}(\hat\ve^0)^{-1}\R
\end{align}
are vectors of the stationary phase of the ordinary and the extraordinary waves,
\begin{equation}
 \label{2.13}
 f_{(o)}=1, \quad f_{(e)}^2=\frac{(\s^{(e)}\hat\ve^0\s^{(e)})
 (\s^{(e)}\hat\ve^{02}\s^{(e)})}{\ve_\parallel\ve_\perp},
 \end{equation}
where the unit vector $\bs^{(j)}={\bf k}^{(j)}/k^{(j)}$ is directed along the wave vector.

The fluctuations of the permittivity tensor are mainly caused by the
fluctuations of the director and have the form
\begin{equation}
 \label{2.2}
\delta\ve_{\alpha\beta}(\r)=\ve_{\alpha\beta}(\r)-\ve_{\alpha\beta}^0
  =\ve_a(n_\alpha^0\delta n_\beta(\r)+
n_\beta^0\delta n_\alpha(\r)),
\end{equation}
where
$\ve_{\alpha\beta}^0=\ve_\perp\delta_{\alpha\beta}+ \ve_an_\alpha^0n_\beta^0$.

In the Born approximation the solution of the integral equation~(\ref{2.8}) yields the
single scattering by fluctuations of the permittivity
$\delta\hat\ve(\r)$. The intensity of the single scattering is equal to
\begin{multline}
 \label{2.14}
 I_{(i)}^{(s)}=I_{(i)}^0\frac{V_{sc}}{(4\pi)^2R^2}
 \frac{1}{n^{(i)}\cos\delta^{(i)}} \\
 \times \sum\limits_{j=o,e}\frac{n^{(j)}}
 {\cos^3\delta^{(j)}}f_{(j)}^2e_\alpha^{(s)}e_\beta^{(s)}
 B_{\alpha\beta\mu\nu}(\q)e_\mu^{(i)}e_\nu^{(i)},
 \end{multline}
\begin{align}
\be^{(o)} &= \dfrac{{\bf n}\times{\bf s}^{(o)}}{\sin\theta_o},\nonumber \\
\be^{(e)} &= \dfrac{\bs^{(e)}\epar\cos \theta_{e}-\bn\left(\epar\cos^2\theta_{e}+\eperp\sin^2\theta_{e}\right)}
                   {\sin \theta_{e}\left(\epar^2\cos^2\theta_{e}+\eperp^2\sin^2\theta_{e}\right)^{1/2}},
\end{align}
where indices $i,s=(o,e)$ show the type of the incident and the scattered waves, $V_{sc}$
is the scattering volume, $R$ is the distance between the scattering volume and the point of observation,
\begin{equation}
 \label{2.15}
\cos\delta^{(o)}=1,\quad \cos\delta^{(e)}=
\frac{(\e^{(e)}\hat\ve^0\e^{(e)})^{1/2}}{n^{(e)}},
\end{equation}
$\q=\k^{(s)}-\k^{(i)}$,  $\k^{(s)}$ and $\k^{(i)}$ are the  wave vectors
of the scattered and the incident wave, $I_{(i)}^0$ is the intensity of the incident light,
$B_{\alpha\beta\mu\nu}(\q)=k_0^4
\langle\delta\ve_{\alpha\mu}\delta\ve_{\nu\beta}^*\rangle(\q)$
is the correlation function of the permittivity fluctuations. It is
determined by the fluctuations of the director according to the expression
\begin{multline}
 \label{2.16}
B_{\alpha\beta\gamma\delta}(\q)=k_0^4
\ve_a^2\sum\limits_{l=1}^2\langle|\delta n_l(\q)|^2\rangle
(a_{l\alpha}a_{l\gamma}n_\beta^0 n_\delta^0 \\ +
a_{l\alpha}a_{l\delta}n_\beta^0 n_\gamma^0+
a_{l\beta}a_{l\delta}n_\alpha^0 n_\gamma^0 +
a_{l\beta}a_{l\gamma}n_\alpha^0 n_\delta^0).
\end{multline}
Here
\begin{equation}
 \label{1.4}
\langle|\delta
n_l(\q)|^2\rangle=\frac{k_BT}{K_{ll}q_\perp^2+K_{33}q_\parallel^2+\chi_a
H^2},\quad l=1,2,
\end{equation}
$${\bf a}_1(\q)=\frac{\q}{q},\,\,\,\, {\bf a}_2(\q)=\n^0\times{\bf a}_1(\q).$$
In the absence of an external field the director fluctuations diverge as $1/q^2$ for $q\to 0$ and the correlation length of fluctuations increases indefinitely. In the presence of an external field these fluctuations are finite and the magnetic coherence length $\xi_H=\sqrt{{K_{33}}/(\chi_a H^2)}$ plays the role of the correlation length.

If we substitute the explicit expression for the correlation function in the formula for intensity of the single light  scattering~(\ref{2.14}), it is easy to see that there is no scattering of the ordinary ray into the ordinary one, (o)$\rightarrow$(o), since the vector director $\n^0$ and  the polarization vector of the ordinary ray are orthogonal. Note that the scattering (o)$\rightarrow$(e) and (e)$\rightarrow$(o) occur for the vector $\q$ with the
length $q\sim k_0|n^{(o)}-n^{(e)}|$. It means that the scattering vector at all angles
remains finite. At the same time the intensity of scattering (e)$\rightarrow$(e)  for small angles has a
sharp peak. The height of this peak increases with decreasing of the field, $H\to 0$. In this case the scattering of the
extraordinary ray into extraordinary one is mainly forward.

If we ignore the intrinsic absorption the extinction coefficient is determined by the
loss of light due to scattering and it has the form~\cite{Stark1997}
\begin{multline}
 \label{2.19}
 \tau_{(j)}(\k^{(j)})=\frac{1}{(4\pi)^2}
 \frac{e_\alpha^{(j)}e_\beta^{(j)}}{n^{(j)}\cos^2\delta^{(j)}} \\ \times
 \sum\limits_{l=o,e} \int d\Omega_\q^{(l)}
 \frac{n^{(l)}e_\nu^{(l)}e_\mu^{(l)}}{\cos^2\delta^{(l)}}
 B_{\alpha\nu\beta\mu}(\k^{(l)}-\q),
\end{multline}
 where $\int d\Omega_\q^{(l)}$ denotes the integration over the surface
 $q=k^{(l)}(\q)$.

Due to the presence  of permittivity fluctuations  it is necessary  while
calculating the average field and  average scattering intensity to
use the Green's function averaged over the inhomogeneities of the medium. This
averaged Green's function in the coordinate representation has the form
\begin{multline}
\label{2.24}
    \langle \hat G^{R/A} \rangle({\bf R}) =\frac{k_0^2}{4\pi R}
   \sum_{j=o,e}n^{(j)}f_{(j)}
    \frac{\e^{(j)}\otimes\e^{(j)}}
    {(\e^{(j)}\hat\ve^0\e^{(j)})} \\ \times \exp\left(\pm i\k_{st}^{(j)}\cdot{\bf R}-
    \frac{R}{2l_{(j)}(\k_{st}^{(j)}/k_{st}^{(j)})}\right).
\end{multline}
where the indices $R$ and $A$ denote the advanced and retarded Green's function,
$\hat G^A=[\hat G^R]^*$, $l_{(j)}=\tau_{(j)}^{-1}$.

When light propagates in rather thick samples of NLC the multiple scattering regime is formed as far as the director fluctuations are not small. In order to describe the intensity of the multiple scattering it is convenient to use
Bethe-Salpeter equation. In the weak scattering approximation  the equation of Bethe-Salpeter
can be solved by iteration. The solution has the form of infinite   series.
The terms of this series correspond to the contributions of different scattering orders.

Usually in this formal solution the sum of the ladder diagrams $\hat L$ is segregated and then the solution
can be written as
\begin{multline}
\label{3.1a}
    \Gamma_{ikjl}(\R_1,\R_4,\r_1,\r_4) =
   \Gamma^0_{ikjl}(\R_1,\R_4,\r_1,\r_4) \\ +
   \int d\R_2 d\R_3 d\r_2 d\r_3
   \Gamma^0_{ikpq}(\R_1,\R_2,\r_1,\r_2) \\ \times L_{pqmn}(\R_2,\R_3,\r_2,\r_3)
    \Gamma^0_{mnjl}(\R_3,\R_4,\r_3,\r_4),
\end{multline}
where
\begin{multline}
\label{3.2}
  \Gamma^0_{ikjl}(\R_1,\R_2,\r_1,\r_2)=
   \langle G^{R}_{ij} \rangle(\R_1+\frac{\r_1}{2},\R_2+\frac{\r_2}{2}) \\ \times
   \langle G^{A}_{lk} \rangle(\R_3+\frac{\r_3}{2},\R_4+\frac{\r_4}{2}),
\end{multline}
\begin{multline}
\label{3.3}
  \Gamma_{ikjl}(\R_1,\R_2,\r_1,\r_2)=
   \langle G^{R}_{ij}(\R_1+\frac{\r_1}{2},\R_2+\frac{\r_2}{2}) \\ \times G^{A}_{lk}(\R_3+\frac{\r_3}{2},\R_4+\frac{\r_4}{2})\rangle,
\end{multline}
The function $\hat \Gamma$ takes into account the  spatial correlations of the Green's function in
inhomogeneous medium. The sum of the ladder diagrams is conveniently presented as
\begin{fmffile}{ladder11}
\begin{equation}
 \label{4.1ladders}
 L_{jikl}=\,
\parbox{4mm}
{\begin{fmfgraph}(4,30)
 \fmftop{t1}
 \fmfbottom{b1}
 \fmf{wiggly}{t1,b1}
 \fmfdot{t1}
 \fmfdot{b1}
 \end{fmfgraph}}
+\;
\parbox{12mm}
{\begin{fmfgraph}(24,30)
 \fmfstraight
 \fmftop{t1,t2}
 \fmfbottom{b1,b2}
 \fmf{plain}{t1,t2}
 \fmf{plain}{b1,b2}
 \fmf{wiggly}{t1,b1}
 \fmf{wiggly}{t2,b2}
 \fmfdotn{t}{2}
 \fmfdotn{b}{2}
\end{fmfgraph}}
+ \;
 \parbox{20mm} {\begin{fmfgraph}(50,30)
 \fmfstraight
 \fmftop{t1,t2,t3}
 \fmfbottom{b1,b2,b3}
 \fmf{plain}{t1,t2,t3}
 \fmf{plain}{b1,b2,b3}
 \fmf{wiggly}{t1,b1}
 \fmf{wiggly}{t2,b2}
 \fmf{wiggly}{t3,b3}
 \fmfdotn{t}{3}
 \fmfdotn{b}{3}
\end{fmfgraph}}
+\ldots \,,
\end{equation}
\end{fmffile}
where the wavy lines represent the correlation functions~$\hat B$, parallel
lines denote the product of averaged  Green's functions~$\hat \Gamma_0$ and each vertex  supposes the
integration over the spatial variable.

Usually  it is more convenient to analyze Bethe - Salpeter equation by passing to
spatial Fourier spectrum. Here we take into account that after
statistical averaging characteristics of the system are  spatially
homogeneous, i.e. they depend on the differences of coordinates only.
Therefore the function $\hat\Gamma$ in~(\ref{3.3}) is the function of only
 three variables  $\R_1-\R_2$, $\r_1$, $\r_2$ out of four. In this case
it is convenient to perform a Fourier transform over all the three variables
\begin{multline}
 \label{3.5}
  \hat\Gamma(\K,\k_1,\k_2)= \int d(\R_1-\R_2) d\r_1 d\r_2 \\
  \exp[-i\K\cdot(\R_1-\R_2)-i\k_1 \cdot \r_1
  +i\k_2 \cdot \r_2] \\ \times\hat\Gamma(\R_1-\R_2,\r_1,\r_2).
\end{multline}
In the Fourier representation the Bethe-Salpeter equation has the form
\begin{multline}
\label{3.5a}
    \Gamma_{ikjl}(\K,\k_1,\k_2) =
   \Gamma^0_{ikmn}(\K,\k_1)I^{(4)}_{mnjl}(\k_1,\k_2) \\ +
   \Gamma^0_{ikpq}(\K,\k_1) \int \frac{d\k}{(2\pi)^3}
   B_{pqmn}(\k_1-\k) \Gamma_{mnjl}(\K,\k,\k_2),
\end{multline}
and the corresponding iterative solution is
\begin{multline}
\label{3.6}
    \Gamma_{ikjl}(\K,\k_1,\k_2) =
   \Gamma^0_{ikmn}(\K,\k_1)I^{(4)}_{mnjl}(\k_1,\k_2) \\ +
   \Gamma^0_{ikpq}(\K,\k_1) \int \frac{d\k}{(2\pi)^3}
   L_{pqmn}(\k_1-\k) \Gamma^0_{mnjl}(\K,\k,\k_2),
\end{multline}
where
\begin{equation}
\label{3.7}
  \Gamma^0_{ikpq}(\K,\k)=
   \langle G^{R}_{ip} \rangle(\k+{\K}/{2})
   \langle G^{A}_{qk} \rangle(\k-{\K}/{2}),
\end{equation}
\begin{equation}
\label{3.7a}
  I^{(4)}_{mnjl}(\k_1,\k_2)=
   (2\pi)^3\delta(\k_1-\k_2)
   \frac{\delta_{mj}\delta_{nl}+\delta_{ml}\delta_{nj}}{2}.
\end{equation}

The approximate solution of the equation~(\ref{3.5a}) was obtained in work~\cite{Stark1997}. This solution corresponds to the diffusion approximation. Difficulty in solving the problem is caused by the fact that it is required to take into account the vector  nature of the electromagnetic field and also different polarizations of eigenwaves of the medium. Therefore the Bethe - Salpeter integral equation was solved in ~\cite{Stark1997} by expansion of the kernel of the integral equation over the eigenfunctions. If in this expansion  we are limited by the minimal eigenvalue the solution of the Bethe - Salpeter equation
has the form~\cite{Stark1997}
\begin{equation}
\label{3.8}
  \Gamma^{(D)}_{ikjl}(\K,\k_1,\k_2)=-\frac1N
  \frac{\Delta G(\k_1)_{ik}\Delta G_{jl}(\k_2)}{(\K\hat D\K)},
\end{equation}
where
$$
N=\frac{2k_0^2}{\pi c}\frac1{8\pi}\sum_{j=o,e}\int d\Omega_\k
n^{(j)3}(\s),
$$
\begin{equation}
\label{3.9}
  \Delta\hat G(\k)=2i\im \langle \hat G^R \rangle(\k),
\end{equation}
$\hat D$ is the tensor diffusion coefficient of light which for NLC has the form
\begin{equation}
\label{3.10}
  \hat D=D_\perp \hat I +(D_\parallel-D_\perp)\n^0\otimes\n^0,
\end{equation}
$D_\parallel$ and $D_\perp$ are the diffusion coefficients along and cross $\n^0$. Tensor diffusion coefficient was obtained in~\cite{Stark1997} by expansion over spherical harmonics taking into account the contribution of the lower modes. The diffusion equation in this case has the form
\begin{equation}
\label{dif1}
 \frac{\partial P}{\partial t}=D_{\parallel}\nabla_\parallel^2 P+D_{\perp}\nabla_\perp^2 P,
 \end{equation}
where $P=P(\r,t)$ is the probability density of arrival for a photon at the point $\r$ at time $t$. In the case of a point source in an infinite medium the solution of this equation is
\begin{equation}
\label{dif2}
 P(\r,t) =\frac{1}{8(\pi t)^{3/2}D_\perp D_\parallel^{1/2}}
         \exp\left[-\frac{1}{4 t}\left(\frac{r_\parallel^2}{D_\parallel}+\frac{r_\perp^2}{D_\perp}\right)\right],
\end{equation}
where $r_\parallel$ and $\r_\perp$ are the directions along and across the director.

The solution of the Bethe-Salpeter equation in the diffusion approximation allows us to describe multiple scattering of light in all directions except  a narrow vicinity of the  backscattering. In this area the effect of the coherent backscattering becomes significant. This effect means that the fields scattered by the same inhomogeneities in reverse order are coherent with fields scattered in the direct order. This leads to additional contribution to the scattering and the emerging of a narrow peak in the backscattering region.

For description of this intensity peak it is required to take into account the sequence of scatterings occurring in the reverse order. Such sequences are described by the cyclic diagrams, $\hat C$,
\begin{multline}
\label{4.3}
  C_{\gamma\delta\mu\nu}(\R_1,\R_2,\r_1,\r_2)
  \\ = \tilde L_{\gamma\nu\mu\delta}
  \left(\frac{\R_1+\R_2}2+\frac{\r_1-\r_2}4,
    \frac{\R_1+\R_2}2-\frac{\r_1-\r_2}4, \right. \\
  \left. \R_1-\R_2+
  \frac{\r_1+\r_2}2,\R_2-\R_1+\frac{\r_1+\r_2}2\right).
\end{multline}
Here, the notation $\tilde L_{\gamma\nu\mu\delta}$ is introduced for the sum of ladder diagrams~(\ref{4.1ladders}) starting with the double-scattering.

{\sloppy We are interested in the intensity of multiply scattered light containing contributions of the ladder and the cyclic diagrams. The Fourier transform of $J_{\alpha\beta}(\R,\k^{(s)})$ in the coherence function
$\langle E_{\alpha}(\R+\r/2)E_{\beta}^*(\R-\r/2)\rangle$ as a function of variable $\r$ is proportional to the intensity of radiation at a point $\R$ with the wave vector $\k^{(s)}$ \cite{Stephen1986},
\begin{equation}
\label{4.30} J_{\alpha\beta}(\R,\k^{(s)})= \int d\r
e^{-i\k^{(s)}\cdot\r} \left\langle E_{\alpha}
\left(\R+\frac{\r}2\right)E_{\beta}^*\left(\R-\frac{\r}2\right)
\right\rangle
\end{equation}
or
\begin{multline}
\label{4.31} J_{\alpha\beta}(\R,\k^{(s)})= \int d\r
e^{-i\k^{(s)}\cdot\r} \int\limits_{V_{sc}} d\R_3 d\r_3 \\
\Gamma_{\alpha\beta\alpha'\beta'}(\R,\R_3,\r,\r_3)
J^0_{\alpha'\beta'}(\R_3,\r_3),
\end{multline}
where $J^0_{\alpha'\beta'}(\R_3,\r_3)$ is the intensity of the source.
We consider the case when the plane wave falling on the scattering volume.
Taking into account  the contributions of ladder and
cyclic diagrams in $\hat \Gamma$, ~(\ref{4.31}), we obtain
\begin{multline}
\label{4.7}
  J_{\alpha\beta}(\R,\k^{(s)})=
  J^{(L)}_{\alpha\beta}(\R,\k^{(s)})+
  J^{(C)}_{\alpha\beta}(\R,\k^{(s)}) \\ =
  \int d\R_1 d\R_2\int\frac{d\k_1}{(2\pi)^3}
  \frac{d\k_2}{(2\pi)^3}
  \Gamma_{\alpha\beta\gamma\delta}^0(\R,\R_1,\k^{(s)},\k_1)
  \\ \times
  [L_{\gamma\delta\mu\nu}(\R_1,\R_2,\k_1,\k_2)+
  C_{\gamma\delta\mu\nu}(\R_1,\R_2,\k_1,\k_2)] \\ \times
  \int\limits_{S_{sc}}d\R_3
  \Gamma_{\mu\nu\alpha'\beta'}^0(\R_2,\R_3,\k_2,\k^{(i)})
  I^{(i)}_{\alpha'\beta'},
\end{multline}
where $I^{(i)}_{\alpha'\beta'}=|E_0^{(i)}|^2e^{(i)}_{\alpha'}e^{(i)}_{\beta'}$, $E_0^{(i)}$ and $e^{(i)}_{\alpha'}$ are the amplitude and the polarization of the incident field, $S_{sc}$ is the area illuminated  by the incident field. In Eq.~(\ref{4.7}) we performed integration over the coordinates $\r$.

}The equation~(\ref{4.7}) is quite general. We are interested in the coherent backscattering in a  medium  which occupies a halfspace.

\begin{figure}[h]
\begin{center}
\includegraphics[width=8cm]{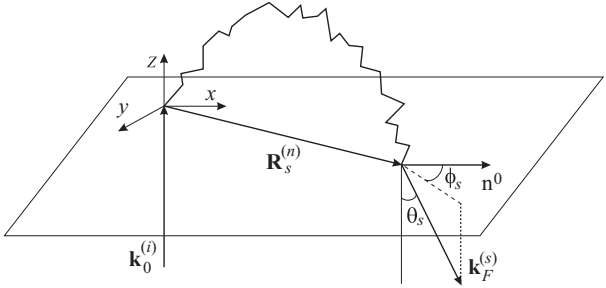}
\end{center}
\caption{Geometry used for calculating coherent backscattering. The scattering medium occupies the half space $z >0$. Here ${\bf R}_s^{(n)}$ is the point of photon emitting from the medium,
$\k_0^{(i)}$ is the wave vectors of the incident wave and $\k_F^{(s)}$ is the wave vector of the photon leaving the medium in the direction determined by the angles $(\theta_s,\phi_s)$ after multiple scattering.}
\label{fig1}
\end{figure}

We consider the geometry presented in Fig.~\ref{fig1} which was used in the experiments~\cite{Wiersma2004,Wiersma2005},   when the NLC is oriented so that the optical axis is parallel to the interface. We assume that the scattering medium occupies the halfspace $z>0$, the optical axis is directed along the $x$ axis, i.e. $\n^0=(1,0,0)$.

Contributions to the intensity of backscattering due to the cyclic and the ladder diagrams
have the form~\cite{AksKuzRom2009}
\begin{multline}
\label{4.25a}
  J_{\alpha\beta}^{(C)}(\R,\k^{(s)})=
  J_{\alpha\beta}^{(C)}(\Theta,\phi) \\ =
  M_{\alpha\beta}^{(C)}(\k^{(i)},\k^{(s)})
  \frac{\pi^2}{8k_0^6}\frac{\ve_\perp}{\ve_\parallel}
  \frac{1}{ D_\perp}\frac{1}{\sqrt{q'^2+1/\xi^2}}
  \\
\times
 \left[\frac{l_{(e)}}
 {\sqrt{q'^2+1/\xi^2}+1/l_{(e)}}
 \frac{\exp(-2z_b\sqrt{q'^2+1/\xi^2})}
 {(\sqrt{q'^2+1/\xi^2}+1/l_{(e)})^2} \right],
\end{multline}
\begin{multline}
\label{4.26}
  J_{\alpha\beta}^{(L)}(\R,\k^{(s)})=
  \frac{\pi^2}{8k_0^6}\frac{\ve_\perp}{\ve_\parallel}
  \left\{M_{\alpha\beta}^{(L)}(\k^{(i)},\k^{(s)})\frac{\xi}{D_\perp} \right.\\ \times
   \left[\frac{l_{(e)}}
 {1/\xi+1/l_{(e)}}-
 \frac{\exp(-2z_b/\xi)}
 {(1/\xi+1/l_{(e)})^2} \right]  + \left.
 \vphantom{\frac{\xi}{D_\perp}}
 M_{\alpha\beta}^{(1)}(\k^{(i)},\k^{(s)})l_{(e)}\right\}.
\end{multline}

Note that in this geometry $\e^{(i)}\|\n^0$ and $\e^{(s)}\|\n^0$. From Eqs.(\ref{2.14}) and~(\ref{2.16})
it is easy to show that there is no  single scattering in this geometry~\cite{Valkov1986}.
Here $z_b=0.71 l$ is the position of the plane where the sum of the ladder diagrams vanishes,
$$
q'^2=k_0^2\left(\frac{D_\parallel}{D_\perp}\Theta^2\cos^2\phi+
\Theta^2\sin^2\phi \right),
$$
\begin{multline}
\label{4.18a}
  M_{\alpha\beta}^{(C)}(\k^{(i)},\k^{(s)}) =
  E^{(e)}_{\alpha\beta\gamma\delta}(\k^{(s)})
 \frac1N  B_{\gamma\nu\mu'\delta'}(\k=0)  g_{\mu'\delta'}
g_{\gamma'\nu'} \\ \times B_{\gamma'\nu'\mu\delta}(\k=0)
E^{(e)}_{\mu\nu\alpha'\beta'}(\k^{(i)}) I_{\alpha'\beta'}^{(i)},
\end{multline}
\begin{multline}
\label{4.18b}
  M_{\alpha\beta}^{(L)}(\k^{(i)},\k^{(s)}) =
  E^{(e)}_{\alpha\beta\gamma\delta}(\k^{(s)})
\frac1N B_{\gamma\delta\mu'\delta'}(\k=0)
g_{\mu'\delta'}g_{\gamma'\nu'}  \\ \times B_{\gamma'\nu'\mu\nu}(\k=0)
 E^{(e)}_{\mu\nu\alpha'\beta'}(\k^{(i)}) I_{\alpha'\beta'}^{(i)},
\end{multline}
$$
 E^{(j)}_{\alpha\beta\gamma\delta}(\k)=e_\alpha^{(j)}(\k)
 e_\beta^{(j)}(\k)e_\gamma^{(j)}(\k)e_\delta^{(j)}(\k),
$$
$$
I^{(i)}_{\alpha'\beta'}
e^{i\k^{(i)}\cdot\r_3}=|E_0^{(i)}|^2e^{(i)}_{\alpha'}
e^{(i)}_{\beta'}e^{i\k^{(i)}\cdot\r_3},
$$
$$
\hat g(\r)=\int \frac{d \k}{(2\pi)^3} e^{i\k\cdot\r}\im\langle
\hat G^{R}\rangle(\k).
$$
While obtaining Eqs.~(\ref{4.25a}) and~(\ref{4.26}) a number of approximations were introduced. In calculating the contribution of ladder and cyclic diagrams, only one type of scattering, $(e)\to(e)$, was taken into account. This approximation is justified by the fact that the scatterings of $(o)\to(e)$ and $(e)\to(o)$ types are much weaker in comparison with $(e)\to(e)$. When calculating the ladder and cyclic diagrams we assumed that the correlation function of the fluctuations of the permittivity tensor $\hat B(\r)$ is short-ranged. However all the polarization factors associated with the liquid crystal were taken into account exactly. And finally the finite thickness of the scattering system $W$ was accounted for as an intrinsic absorption of the scattered radiation with a characteristic decay length $\xi\approx W$. A similar approach was developed in~\cite{mackintosh1988}.

\section{Simulation of scattering in anisotropic medium}\label{sec3}

We have simulated the multiple scattering by Monte Carlo method. In our study of the photon diffusion  we assumed that the entire space is occupied by NLC oriented by the magnetic field. In this case such an approach is justified since the tensor
diffusion~(\ref{3.10}) is a macroscopic quantity and it is independent of the size of the sample. When studying  the coherent backscattering it was considered that NLC fills a halfspace.

The standard   simulation  procedure of the multiple scattering is as follows. Between scatterings the photon propagates along a straight line. The length passed by a photon between successive scatterings is generated randomly so that the average distance between successive scatterings coincides with the mean free path of a photon $l_{(j)}=\tau_{(j)}^{-1}\left(\ki\right)$. Choice of the direction for the photon propagation after scattering is also a random quantity and depends on the single-scattering phase function.

\begin{figure}[h]
\begin{center}
\includegraphics[width=8cm]{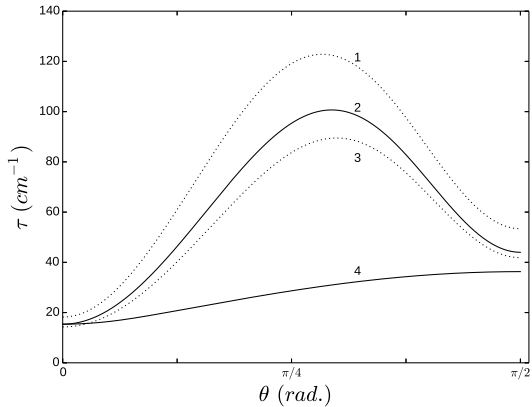}
\end{center}
\caption{Angular dependence of extinction of the ordinary (curve 4) and extraordinary
(curves 1-3) rays. Calculations are carried out by Eq.~(\ref{2.19}) for the
following values of the parameters $\varepsilon_\perp=2.2$, $\varepsilon_a=0.8$,
$\lambda=4.88\cdot10^{-5}$~cm, $T=301$~K, $H=5000$~Oe,
$\chi_a=1.38\cdot 10^{-7}$.
The Frank moduli are as follows: $K_{11}=0.79\, K_{33}$, $K_{22}=0.43\,K_{33}$,
$K_{33}=6.1\cdot 10^{-7}$~dyn for curves 2 and 4;
$K_{11}=2.6\cdot 10^{-7}$~dyn, $K_{22}=1.4\cdot 10^{-7}$ dyn for
curve 1; and $K_{11}=K_{22}=K_{33}=4.5\cdot 10^{-7}$ dyn for curve 3
(one constant approximation). In all three cases, the sum of the Frank moduli is
the same.}
\label{fig2}
\end{figure}

In our problem this procedure is much more complicated. As long as the medium is uniaxial  there exist two types of waves, i.e. ordinary and extraordinary in the scattering process. The value of the wave vector of extraordinary wave depends on the angle between the direction of the wave propagation and that of the vector director. The phase function of the single scattering of this wave depends on the angle between the direction of the optical axis and the wave vectors  of the incident and of the scattered waves. Finally, the mean free path of a photon $l_{(j)}$ determined by the extinction depends on the type of the wave and of the direction of its propagation. The angular dependence of the extinction coefficient in~(\ref{2.19}) determines both the LC optical anisotropy and the modules of orientation elasticity. This dependence does not disappear even if we neglect the difference between the Frank modules~\cite{HeiderichMaynard1997,KuzminValkov2011}. Figure~\ref{fig2} shows the angular dependence of extinction for different ratios between the Frank modules  and the same optical anisotropy.

According to our model prior to each act of scattering a photon is in one of  two ``channels'' of scattering i.e. it has one of  two polarizations, $(o)$ or $(e)$. If prior to scattering a photon had $(e)$ polarization it can be scattered into one of two channels, $(o)$ or $(e)$. At the start of simulations we calculate the probability of $(e)\to(e)$ scattering. This probability $w(\theta_i)$ is determined  as a ratio of the scattering intensity in the extraordinary wave and the total scattering intensity.
\begin{equation}
w(\theta_i) = \frac{\int I_{(e)}^{(e)} d\Omega_{\q}}{\int \left(I_{(e)}^{(e)} + I_{(e)}^{(o)} \right) d\Omega_{\q}},
\end{equation}
where $\theta_i$ is the angle between $\k^{(i)}$ and $\n_0$.

Between successive scatterings the photons move in a medium with permittivity $\varepsilon(\k)$. At each scattering event it is necessary determined randomly the direction of the photon propagation, the type of the wave into which it is supposed to scatter, and the distance of its free path.

The probability density of the scattering  per unit  solid angle is given by the phase function of the single scattering. As far as the investigated system is anisotropic this probability depends on the angle $\theta_i$ between the wave vector $\ki$ prior to the scattering and the vector director ${\bf n}$.

Usually the generation of a random variable with a given probability density is performed using the inverse function method~\cite{sobol}. Typically in this approach a model phase function (for example the Henyey-Greenstein phase function) is used~\cite{Isimaru,Scipetrov1998,Kuzmin2004}. In our case expression for the phase function~(\ref{2.14}) is much more complicated. Therefore generation of  random directions in each scattering event leads to cumbersome procedure including the conversion of special functions such as elliptic integrals. For this reason this approach seems to be hopeless for numerical simulation.

In the general case in order to describe the single scattering it is necessary to define four parameters: two angle determining the direction of the vector $\ki$ and the two angles determining the vector $\ks$. As far as our system is optically uniaxial there is a symmetry consisting in that the simultaneous rotation vectors $\ki$ and $\ks$ around the director $\n$ does not change the scattering probability for $\ki$ to $\ks$. This allows to reduce the number of parameters from four to three. As one of the parameters we choose the angle $\theta_i$. For a given vector $\ki$ we introduce a local Cartesian coordinate system with unit vectors
\begin{equation}\label{200}
    \bv_3=\frac{\ki}{k^{(i)}},\quad \bv_1=\bv_3\times\n,\quad \bv_2=\bv_3\times\bv_1.
\end{equation}
Within this coordinate system the corresponding spherical coordinate system is introduced. The angles $\theta$ and $\phi$ of this coordinate frame   will be used for description of the scattering direction. Thus in order to generate a random direction of the photon after the scattering  it is necessary for a given angle $\theta_i$ to randomly get the angles $\theta$ and $\phi$ defining the direction of $\ks$. For the angle $\phi$ it is sufficient to consider the interval $[0,\pi]$ since
expression~(\ref{2.14}) has a mirror symmetry relative to the point $\phi=\pi$.

Owing to the complexity of the expression~(\ref{2.14}) we have built an approximation of the phase function. For this purpose we took into account that it smoothly depends on the angle $\theta_i$. In order to interpolate the phase function~(\ref{2.14}) we  create initially a discrete set of angles
$\{\tilde\theta_1,\tilde\theta_2,...\tilde\theta_M\}$ for the angle $\theta_i$, such that within the intervals
$\tilde\theta_j \leq \theta_i \leq \tilde\theta_{j+1}$ the phase function changes by less than $1\%$. For each angle
$\tilde\theta_j$ in this set we divide the solid angle of scattering $\theta\in[0,\pi]$, $\phi\in[0,\pi]$ into rectangular cells $s^j$
\begin{equation}
 \label{5.2}
s^j:\quad\theta\in\left[\theta^j_l,\theta^j_r\right),\quad\phi\in\left[\phi^j_l,\phi^j_r\right),
\end{equation}
where the indices $l$ and $r$ refer to the left and the right edges of the rectangle. The sizes of the rectangles are chosen so that the bilinear interpolation of the phase function $I^j(\theta,\phi)$ constructed from the values calculated at the vertices of the cell described the phase function with the required accuracy. The division was carried out adaptively so that a narrow peak of the phase function had a sufficiently large number of cells because photons are scattered primarily in this direction. The construction of interpolation was performed separately for each type of scattering.

In our simulations the accuracy of the interpolation was about $1\%$. This precision was achieved for the number of cells $3\cdot10^3 - 3\cdot10^4$.

In  simulating each act of a single scattering for the angle $\theta_i$ we select the closest element of the set $\{\tilde\theta_1,\tilde\theta_1,...\tilde\theta_M\}$ and this way we get the corresponding interpolation of $I^j(\theta,\phi|\theta_i)$.

The probability that the direction of scattering enters  the cell $s^j$ is equal to
\begin{equation}
\label{5.3}
    p_j=\frac{\int\limits_{\phi,\theta\in s^j}I^j(\theta,\phi|\theta_i)d\Omega}{\sum\limits_t\int\limits_{\phi,\theta\in s^t}I^t(\theta,\phi|\theta_i)d\Omega}.
\end{equation}
With the aid of the uniform random number generator in
 the interval $[0,1]$ we select the cell with the number $t$ of the array such that
\begin{equation}
\label{d0}
\sum_{j=1,j<t}{p_j} < r,\quad\sum_{j=1,j\le t}{p_j} \ge r,
\end{equation}
where  $r$ is a random number.
After selection of the cell it is necessary to choose random  angles $\theta$ and $\phi$ inside the cell. This can be done by using a two-dimensional analogue of the inverse function method~\cite{sobol} using the corresponding bilinear interpolation of the cell $I^t$ as the probability density. Thus the obtained angles $\theta$ and $\phi$ belonging to this cell will determine the direction of the photon propagation after the last scattering.

On having determined  the wave vector of the scattered photon $\ks$ it is necessary to get the distance that the photon  passes before the next scattering event. The probability density distribution of the mean free path between  two successive scattering events $s$ has the form~\cite{sobol}
\begin{equation}
\label{d1}
f(s)=\frac1 l \exp(-s/l),
\end{equation}
where $l$ is the mean free path. Then the probability that the mean free path of a photon exceeds $s$ is
\begin{equation}
\label{d2}
\xi=\int_{s}^{\infty}{f(s')ds'},
\end{equation}
where $\xi$ is a random number uniformly distributed within the range $(0,1]$.  So explicit expression for  $s$ is obtained from Eqs.~(\ref{d1}), (\ref{d2})
\begin{equation}
\label{d3}
s=-{l}{\ln{\xi}}.
\end{equation}
The expression~(\ref{d3}) allows to receive the lengths of the free path of photons with the given   probability density ~(\ref{d1}).

So in our simulation we first choose the scattering channel, then the direction of scattering, and finally the distance that the photon would run before the next scattering.

\section{Radiative transfer in the diffusion approximation}

In the studying of the photon diffusion the purpose of  simulation is to investigate the statistical features of radiation transfer in an anisotropic medium. For this purpose we run randomly a separate photon and determine its trajectory by the method described in the previous section. The procedure is repeated many times in order to  get a set of trajectories. With the help of this set we calculate the mean square displacement of photons along and across to the director, $\langle r_\parallel^2\rangle$ and $\langle \r_\perp^2\rangle$. It is known that starting from a certain moment of time the scattered radiation can be described within the framework of the diffusion approximation. It means that the mean square displacement of photons starts to depend linearly on time. In this case the following relations are valid
\begin{align}
    \label{4.1}
                   \langle x^2\rangle = & \langle r_\parallel^2\rangle = 2D_\parallel t \\
    \label{4.2}
\langle y^2\rangle + \langle z^2\rangle = & \langle \r_\perp^2\rangle = 4D_\perp t.
\end{align}
Here the direction of $x$ axis is chosen along the director. The performed calculation allows to study the transition to the diffusion regime with increasing of the scattering order. Also we investigate the dependence of the diffusion coefficients on the external field and on the wavelength of light~\cite{prog1}.

\begin{figure}[h]
\begin{center}
\includegraphics[width=8cm]{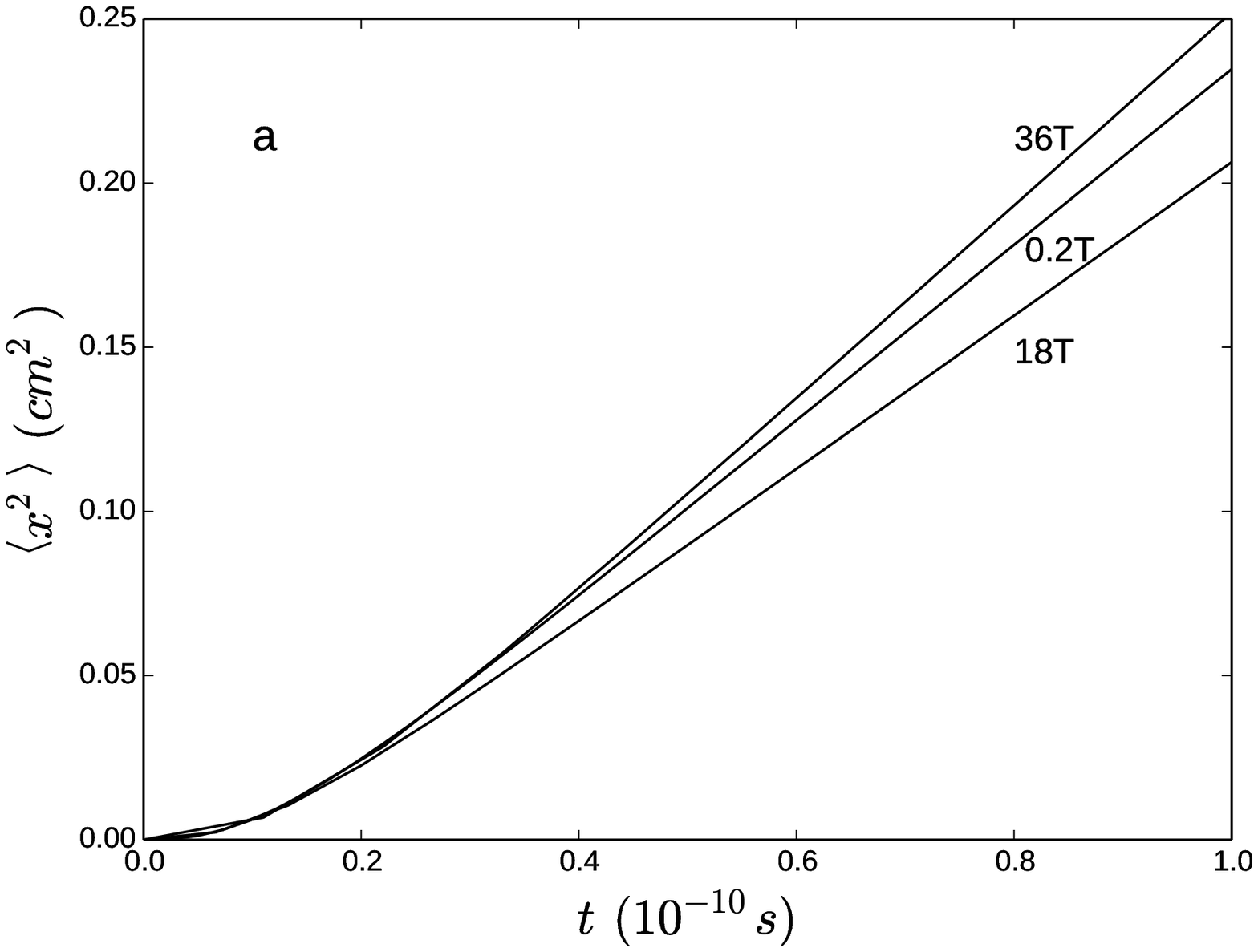}
\includegraphics[width=8cm]{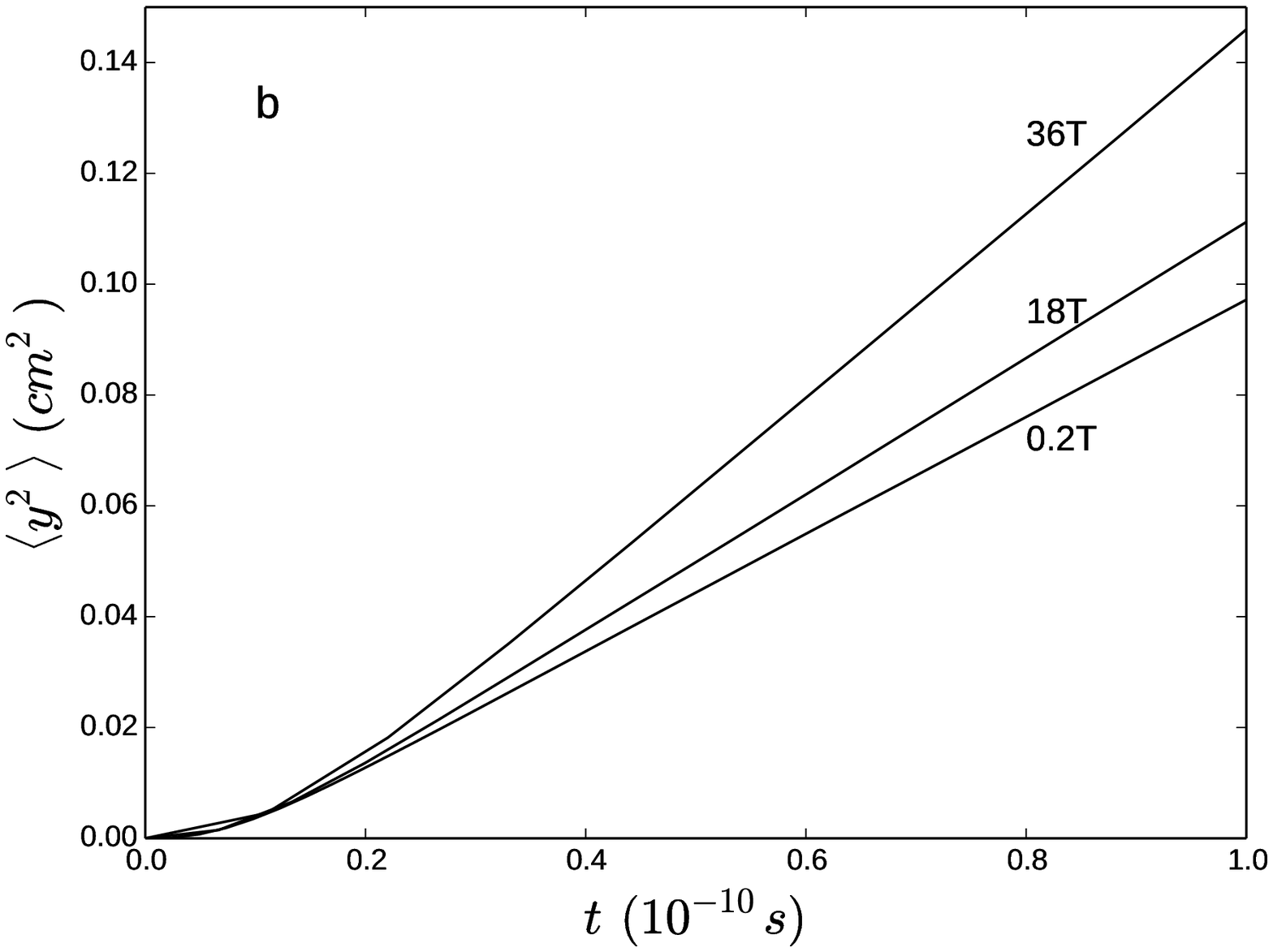}
\end{center}
\caption{Plots of the mean square displacement of the photons (a) along and (b)
across the director vs. time for three values of the external magnetic field.}
\label{fig3}
\end{figure}

\begin{figure}[h]
\begin{center}
\includegraphics[width=8cm]{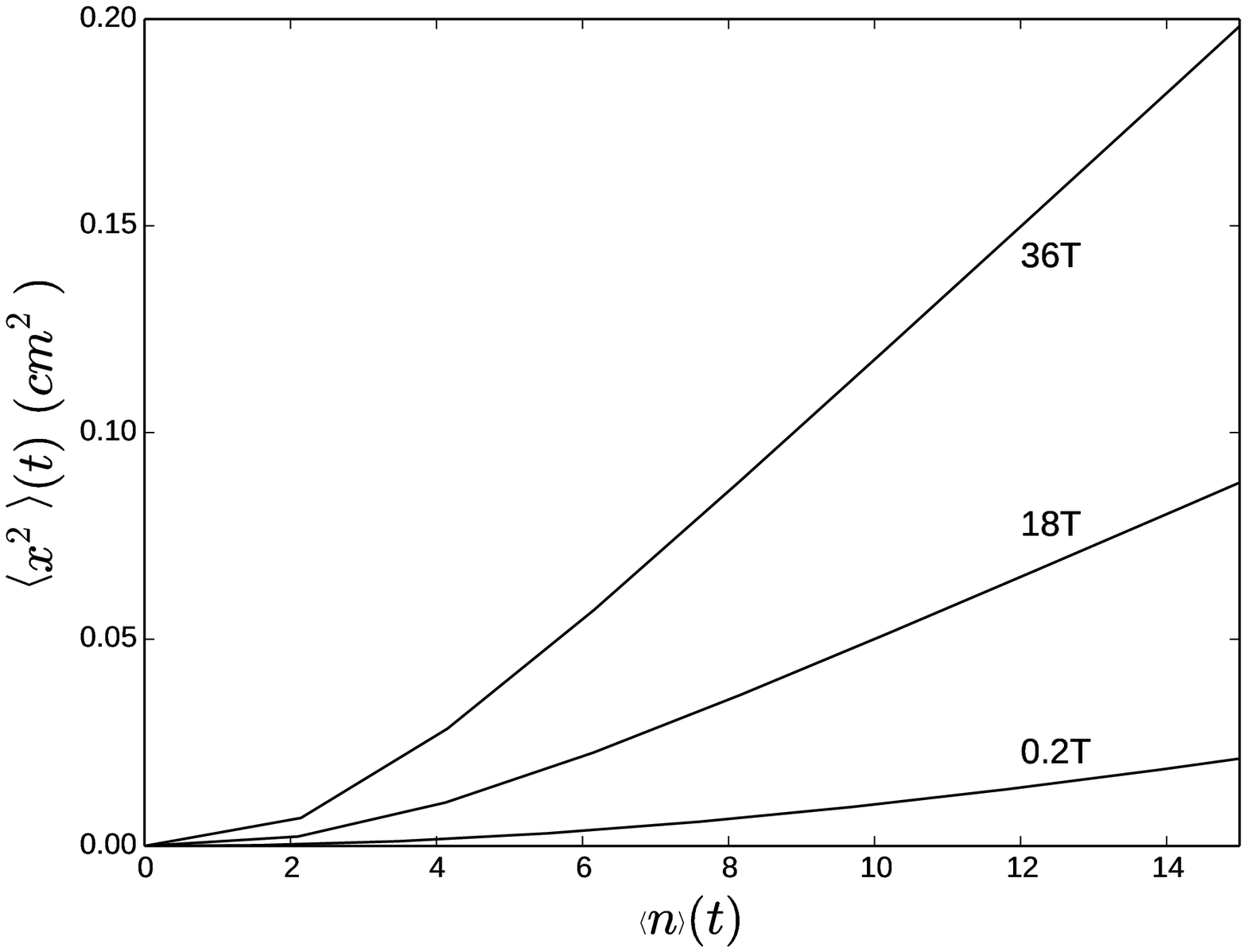}
\includegraphics[width=8cm]{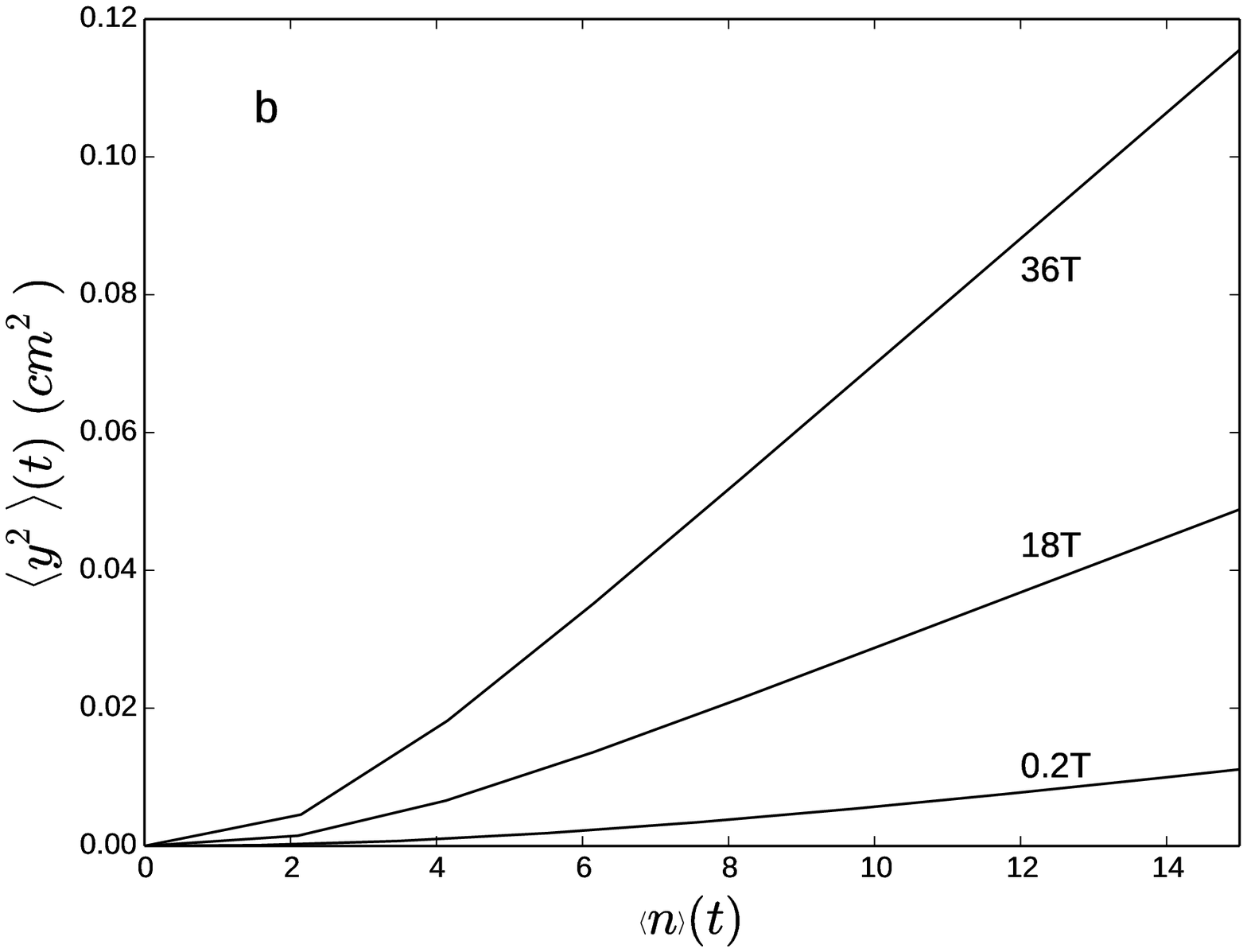}
\end{center}
\caption{Plots of the mean square displacement of the photons (a) along
and (b) across the director vs. the mean number of
scattering events for three values of external magnetic fields.
The plots were obtained in the following way. The number of scatterings, $n$, and the square of photon coordinates, $x^2$ and $y^2$, were calculated for each trajectory and for each moment of time. These values were averaged over the trajectories and $\langle n \rangle (t)$, $\langle x^2 \rangle (t)$, $\langle y^2 \rangle (t)$  were obtained for each $t$.
}
\label{fig4}
\end{figure}

Figure~\ref{fig3} shows the time dependence of the mean squared displacement of the photon along, 1a, and  across, 1b,  the director  for three values of external magnetic field. In Fig.~\ref{fig4} the  same value as a function of the average scattering orders is presented.
The interval of linear dependence in both figures correspond to areas where the diffusion approximation is valid. In the simulation process all photons are emitted in the same direction which is taken for $z$ axis.

It is seen that the initial part of the square displacement of photons does not linearly depends on time. It means that in this interval the diffusion regime has not yet established. Note that the forms of the curves in Figs.~\ref{fig3} and~\ref{fig4} do not coincide. The reason is that the transition from the time scale to the scale of scattering orders is rather complicated. The reason is that the passage time between successive scatterings of photons depends on the direction of propagation. For the extraordinary beam it is due to  dependence of the refractive index and the extinction coefficient on the angle between the wave vector and the director. For an ordinary beam it is caused by the angular dependence of the extinction coefficient only.

It is seen that the number of scattering orders which are required for transition to the diffusion regime decreases with increasing of magnetic field. This result is natural since with increasing of magnetic field the phase function of the single scattering approaches to circular one and consequently the randomization of the photon propagation directions occurs for a smaller number of scatterings.

\begin{figure}[h]
\begin{center}
\includegraphics[width=8cm]{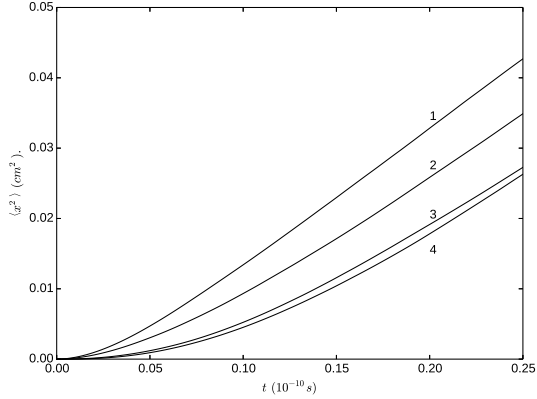}
\end{center}
\caption{Dependence of the mean square displacement of the photon,  $\langle x^2 \rangle$, on angle
$\theta$ between the incident beam and the director: (1) -- $\theta = \pi/4$, (2) -- $\theta =  \pi/3$,
(3) -- $\theta = \pi/12$, and (4) -- $\theta = \pi/2$.
}
\label{fig5}
\end{figure}

We also studied the dependence   of the  transition rate to the diffusion regime on the direction of photon  emission  from the source. It was found that the time of transition to diffusive regime is much smaller if the photon is emitted within the angular range $40-60^\circ$ with respect to the vector director. This is probably due to the fact that in this
interval of angles the photon free path length  as it is shown in Fig.~\ref{fig2}, is minimal. Consequently during the same time a photon experiences greater number of scatterings. In order to illustrate this effect we  present functions $\langle x^2\rangle(t)$ for the four directions of emission of photons in Fig.~\ref{fig5}.

The diffusion coefficients were found from the array of data corresponding to the diffusive regime by the method of the maximum likelihood using the explicit form of the Green function~(\ref{dif2}) of the diffusion equation. They were found from the relations
\begin{equation}
\label{101}
D_\perp=\frac{1}{4MN}\sum\limits_{j=1}^M\frac{1}{t_j}\sum\limits_{l=1}^N {r^2}_{\perp l}(t_j) =\frac{1}{4M}\sum\limits_{j=1}^M\frac{1}{t_j}
\langle {r^2}_{\perp}(t_j)\rangle,
\end{equation}
\begin{equation}
\label{102}
D_\parallel=\frac{1}{2MN}\sum\limits_{j=1}^M\frac{1}{t_j}\sum\limits_{l=1}^N {r^2}_{\parallel l}(t_j) =\frac{1}{2M}\sum\limits_{j=1}^M\frac{1}{t_j}
\langle {r^2}_{\parallel}(t_j)\rangle,
\end{equation}
where $N$ is the number of  radiated photons, $M$ is the number of time steps, the index $l$ is the  number of photons, and the index $j$ is the time step. In our calculation the discrete time  points  $t_j$ are introduced starting from $t = t_0$ with a constant step $\Delta t$. Here time $t = t_0$ belongs to the region of the diffusion regime.
In general we can choose a rather arbitrary set of values of $t_j$ as long as for such separate trajectories it is not difficult to  determine the position of the photon at each moment.

\begin{figure}[h]
\begin{center}
\includegraphics[width=8cm]{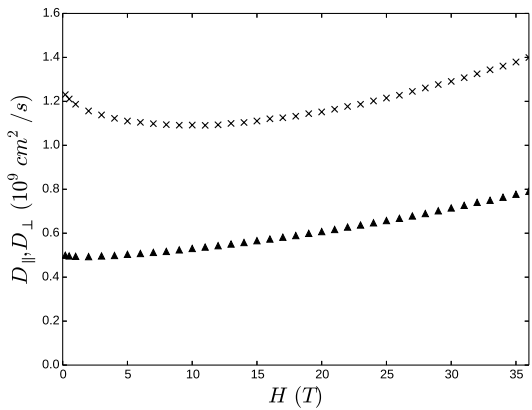}
\end{center}
\caption{Dependence of the diffusion coefficients of photons
$D_\parallel$ ($\times$) and  $D_\perp$ ($\blacktriangle$)  on the magnetic field.
}
\label{fig6}
\end{figure}

\begin{figure}[h]
\begin{center}
\includegraphics[width=8cm]{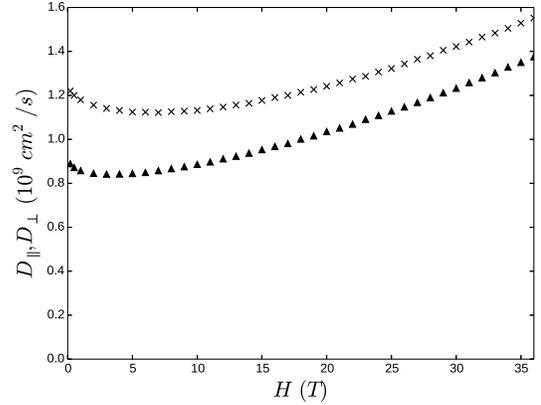}
\end{center}
\caption{Dependence of the diffusion coefficients of photons
$D_\parallel$ ($\times$) and  $D_\perp$ ($\blacktriangle$)  on the magnetic field calculated with
regard to the (e) rays only. }
\label{fig7}
\end{figure}

\begin{figure}[h]
\begin{center}
\includegraphics[width=8cm]{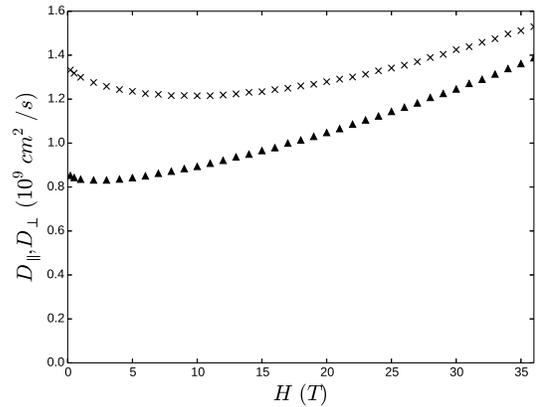}
\end{center}
\caption{Dependence of the diffusion coefficients of photons
$D_\parallel$ ($\times$) and  $D_\perp$ ($\blacktriangle$)  on the magnetic field calculated with
regard to the (e) rays only. Simulations are performed for
one constant approximation ($K_{11}=K_{22}=K_{33}$).
}
\label{fig8}
\end{figure}

Figure~\ref{fig6} shows the dependence of the diffusion coefficients on the magnetic field. The figure shows that the diffusion coefficients  vary nonmonotonically with  increase of the field.
One can see that non-monotonic behavior disappears for the strong fields. The reason is that in this case the phase function of the single scattering becomes close to the circular one and the director fluctuations are suppressed by the external field.
In order to determine the cause of the non-monotonic behavior we have tried to simplify the system which was used in the simulation. Figure~\ref{fig7} shows the results of the calculation when only $(e)\to (e)$ scattering is taken into account. The results   when in addition one-constant approximation is used are shown in Fig.~\ref{fig8}. It is seen that the non-monotonic dependence of the light diffusion coefficients  on the magnetic field holds.
The observed non-monotonic behavior of the diffusion coefficient does not agree with the results of calculation in~\cite{StarkKao1997,Stark1997}, where smooth growth of $D_\parallel$ and $D_\perp$  with increasing of the field is predicted. Perhaps such a discrepancy is due to the fact that in~\cite{StarkKao1997,Stark1997} it was taken into account only the minimal  eigenvalue of the integral operator of the Bethe-Salpeter equation. As far as the account for the following terms of the series is a rather complicated problem we have illustrated the possible reason of a non-monotonic dependence on the field for a simple model system.

We will consider a scalar isotropic model for which the diffusion coefficient has the form
\begin{equation}
\label{1000}
D=\frac{v}{3}\frac{1}{(\mu'_s+\mu_a)}=\frac{v}{3}\frac{1}{\left(\frac{1-\langle\cos\theta\rangle}{l_{ext}}+
\frac{1}{l_a}\right)},
\end{equation}
where $l_a$ is the absorption length, $l_{ext}$ is the extinction length, $\langle\cos\theta\rangle$  is the average cosine of the scattering angle. For the phase function of the single scattering we accept
\begin{equation}
\label{1001}
I=\frac{1}{N}\frac{1}{Kq^2+\chi_a H^2},
\end{equation}
where $N$ is the normalization coefficient, $q=2k\sin({\theta}/{2})$, $\theta$ is the scattering angle. Actually in the framework of this model it is possible to describe the multiple scattering of light in the critical region if we use the Ornstein-Zernike approximation. In this case  the correlation length of fluctuations of the order parameter is used instead of magnetic coherence length $\sqrt{{K}/({\chi_a H^2})}$. For this model, the diffusion coefficient has the form
\begin{equation}
\label{1005}
D=\frac{v}{3}\frac{1}{2/\{l_{ext}[\ln(1+2/h^2)-h^2]\}+{1}/{l_a}},
\end{equation}
where $h^2={\chi_a H^2}/(2k^2K)$,
\begin{equation}
\label{1004}
l_{ext}=\frac{1}{2(4\pi k)^2KN}\frac{1}{\ln\left(1+{2}/{h^2}\right)}.
\end{equation}

\begin{figure}[h]
\begin{center}
\includegraphics[width=8cm]{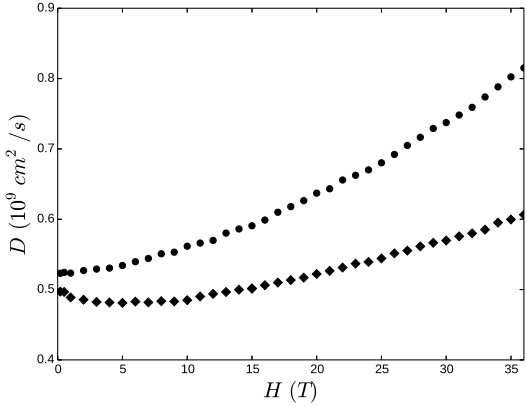}
\end{center}
\caption{Dependence of the diffusion coefficient on the external magnetic field for the
scalar model~(\ref{1001}) without ($\bullet$) and with ($\blacklozenge$) self-absorption.
}
\label{fig9}
\end{figure}

For this model we calculate  the dependence of the diffusion coefficient on the field using Eq.~(\ref{1005}) and also using  the program that was created earlier for simulation of the multiple scattering of light in NLC. The results  coincide within  fractions of  percent. Figure~\ref{fig9} shows the dependence of $D$ on $H$, it is seen that $D$ indeed may increase nonmonotonically with the growth of $H$. Such behavior is caused by presence  an additional length of absorption $l_a$. If we remove the absorption coefficient, $\mu_a=0$, the nonmonotonic behavior disappears.

Monotonic increase of the diffusion coefficient with the growth of the field is explained by the suppression of director fluctuations when the field is being increased. This leads to the increase of the extinction length and the value $1-\langle\cos\theta\rangle$ in Eq.~(\ref{1000}). However the extinction length increases slightly faster. Account of the absorption violates this synchronization and it leads to non-monotonic behavior of the diffusion coefficient.

In the case of NLC the reason of non-monotonic behavior is more complicated. It is probably that the length of extinction and the phase function of the single scattering depend on the direction of the photon propagation.

\begin{center}
\begin{table}
\caption{\newline Comparison of the calculated and measured diffusion
coefficients of photons}
    \begin{tabular}{| l | l | l | l | l |}
    \hline
    No.& Results of & $D_\parallel, 10^9 \text{cm}^2/\text{s}$ & $D_\perp,
    10^9 \text{cm}^2/\text{s}$ & $D_\parallel / D_\perp$ \\ \hline
    1& Calculations& 1.2 & 0.50 & 2.46 \\ \hline
    2& Experiments~\cite{StarkKao1997} & 0.7 $\pm$ 0.1 & 0.5 $\pm$ 0.1 & 1.6 $\pm$ 0.25 \\ \hline
    3& Calculations& 0.61 & 0.25 & 2.45 \\   \hline
    4& Experiments~\cite{Wiersma2004, Wiersma2005} & $0.456 \pm 0.019$ & $0.362 \pm 0.015$ & 1.26 \\ \hline
    \end{tabular}
\end{table}
\end{center}

We also compared the obtained results with the available experimental data.
In our model we did not use the simplifying assumptions concerning the properties of the liquid crystal such as the one-constant approximation. It allowed us to perform the direct comparison with the experimental
data~\cite{Wiersma2004,Wiersma2005,StarkKao1997}. Our calculations performed for comparison with the experiments were carried out at temperatures, light wavelength and the magnetic field exactly the same as in  the experiments ~\cite{Wiersma2004,Wiersma2005,StarkKao1997}. In these papers measurements were carried out on the liquid crystal 5CB. In this liquid crystal the phase transition isotropic-nematic phase takes place at the temperature $T_c=35.1^\circ$C.  Near $T_c$  parameters of the liquid crystal are very sensitive to the  temperature. In  works~\cite{Wiersma2004,Wiersma2005} the measurements were carried out at the temperature $T=27^\circ$C, the magnetic field $H=0.5$~T and the light wavelength $\lambda=405$~nm. At this temperature Frank modules are  equal  $K_{33}=6.1\times 10^{-7}$~dyne, $K_{11}=0.79K_{33}$~dyne, $K_{22}=0.43K_{33}$~dyne~\cite{Wiersma2004,Wiersma2005,Pine1990}. In~\cite{StarkKao1997} measurements were made at $T=30^\circ$C, magnetic field $H=0.2$~T and the light wavelength $\lambda=514.5$~nm. In our we have taken $K_{33}=7.5\times 10^{-7}$~dyne~\cite{Bradshaw1985} and the relation between Frank modules were taken the same as at  temperature $T=27^\circ$C. The results of simulation and the corresponding  experimental data are shown in the Table. Here lines~1 and~3 are the results of computer simulation with the parameters used in the experiments~\cite{StarkKao1997} and~\cite{Wiersma2004,Wiersma2005}. Lines~2 and~4 are the experimental values. The Table shows that the absolute values of the diffusion coefficients are in good agreement in the case of $D_\perp$  but for $D_\|$ simulation predicts higher values compared with the experimental ones.

\begin{figure}[h]
\begin{center}
\includegraphics[width=8cm]{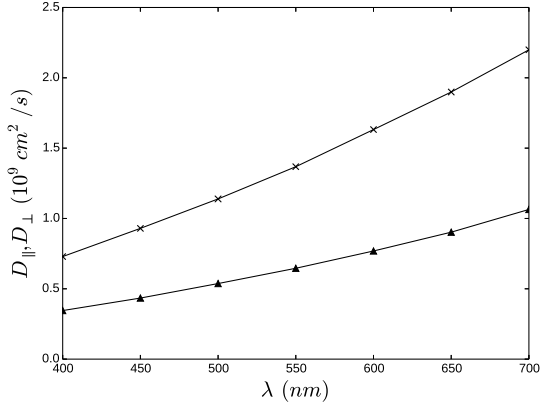}
\end{center}
\caption{Dependence of the diffusion coefficients of photons
$D_\parallel$ ($\times$) and  $D_\perp$ ($\blacktriangle$)  on the  radiation wavelength.
}
\label{fig10}
\end{figure}

We also investigated the dependence of the diffusion coefficients on the wavelength of  light. The results of the calculations are shown in Fig.~\ref{fig10}. It is seen that within the visible light spectrum the diffusion coefficients varies almost twice. This means that in quantitative calculations this effect can be essential.

In the analysis of the   diffusion tensor dependence on the magnetic field  calculations were carried out up to the fields when the values of the orientational and of the field contributions to the energy are comparable. It corresponds to an anomalously high values of the field. The reason is that all estimates refer to the diamagnetic liquid crystals which have very low anisotropy of magnetic susceptibility. At present paramagnetic liquid crystals based on rare-earth elements are synthesized in which the anisotropy of magnetic susceptibility has  much greater values. For these liquid crystals even a strong field region is quite available for experimental research.

\section{The formation of the coherent backscattering peak}\label{sec4}

The main feature of calculating the coherent backscattering peak in nematic liquid crystals is that noticeable interest in  scattering of photons emitted from the medium presents the scattering extremely close to the normal, i.e. at  angles
$\sim 10^{-4}\div10^{-6}$~rad. After multiple scattering the probability of photon emission  in such a narrow range of angles is very small. Therefore  the direct numerical procedure based on a simple counting of photons in this range of angles is extremely inefficient, and in order to solve this problem a semi-analytical Monte Carlo~\cite{Tinet1996} method is used~\cite{prog2}.
The idea of this method is as follows. We take into account the contribution $\delta_n(\ks_F)$  of each photon  into intensity at each scattering event in the direction $\ks_F$ in the range of angles $\theta_s < 10^{-4}$ rad that we are interested in
\begin{equation}
\label{6.1}
    \delta_n(\ks_F)  =W_n p_{is}(\ki,\ks_F)\exp\left[-\frac{1}{l_s(\ks_F)}\frac{z_n}{\cos\theta_s}\right],
\end{equation}
where $n$ is the scattering order, $W_n$ is the weight of a photon, $p_{is}$ is the  normalized single scattering phase function, $\ks_F$ is the wave vector of scattering directed to the receiver at an angle $\theta_s$, $z_n$ is the distance between the current position of the photon and the boundary. The angle $\theta_s$ is measured strictly from the backscattering direction  $0\leq\theta_s\leq{\pi}/{2}$. Equation~(\ref{6.1}) has a simple physical meaning.
The contribution of the photon $\delta_n(\ks_F)$ is the product of three factors: the probability of a photon $W_n$ to suffer $n$ scatterings without abandoning the medium; the probability $p_{is}$  to have the direction of the scattering $\ks_F$;  the exponential factor meaning the probability to reach the boundary without experiencing  collisions. It should also be taken into account that a photon  contributes into $(o)$ or $(e)$ scattering canal.

Calculations have shown that in our system the photons are emitted from the medium mainly after a small number of scatterings, $n\sim 10^2$. On the other hand a significant contribution to the coherent backscattering yield the photons that have experienced a very large number of scatterings $n\sim 10^4\div10^5$. For better accounting  for these photons we used a modified procedure of simulation in which the photons do not abandon the medium~\cite{Tinet1996}. In this condition we take into account the weakening of the intensity caused by the weight $W_n$. The condition of keeping  the photons in the medium is reached in the following way. If the current scattering  is directed to the boundary, $k^{(s)}_z < 0$, then the length of the free path is generated between $0$ and the distance from the current  starting point and the boundary.

For the photon located at a distance $z$ from the boundary and having a wave vector $\ki$ before the scattering  the probability to leave the medium at each scattering event is expressed as
\begin{multline}
\label{weight}
    \esc\left(\ki,z\right)=\sum_{s=o,e}\int_0^{{\pi}/{2}}\sin\theta_s d\theta_s\int_{0}^{2\pi}d\phi_s p_{is}\left(\ki,\ks\right) \\ \times \exp\left[-\frac{1}{l_s\left(\ks\right)}\frac{z}{\cos\theta_s}\right],
\end{multline}
where $\phi_s$ is the azimuthal angle measured from the axis $x$.  It is convenient to calculate the function $\esc\left(\ki,z\right)$ preliminarily in the form of an interpolation table. This function provides  determination of the decrease in weight for the photon after each next step of scattering
\begin{equation} \label{6.4}
    W_{n+1}=W_n\left(1-\esc\left(\ki,z\right)\right),\,\,W_1=1.
\end{equation}
The expression~(\ref{6.4}) takes into account the loss of intensity due to emission of photons from the medium.

We assume that the detector that collects the radiation is infinite and occupies the whole plane $XY$. We are interested in the distribution of the photons emitted from the medium  over angles $\theta_s$ and $\phi_s$. For each emitted photon we collect the emission angles $\theta_s$, $\phi_s$ and vector ${\bf R}_s^{(n)}$ which indicate the place of the photon emission from the medium. Vector ${\bf R}_s^{(n)}$ lies on the surface of the medium, $R_{s,z}^{(n)}=0$.
Summation of the contributions of photons $\delta_n(\ks_F)$ (\ref{6.1}) in the direction $\theta_s$, $\phi_s$  yields the angular distribution of the intensity in the ladder approximation. The corresponding contribution of the cyclic diagrams assumed to be obtained by multiplying $\delta_n(\ks_F)$ by the phase factor $\cos[{\bf q}\cdot({\bf R}_s^{(n)}-{\bf R}_i)]$ \cite{Maret1987,Pine1990}, where  vector ${\bf R}_i$ indicates the place of the photons incidence, its coordinates are taken ${\bf R}_i=(0,0,0)$,
 ${\bf q}$ is the scattering vector ${\bf q}=\ks_F-\ki_0$ and ${\bf R}_i$.
Summing over all the photons for each pair of angles $\theta_s$ and $\phi_s$ we get the angular dependence of the relative intensity of the scattered radiation~\cite{mackintosh1988,Stephen1986}
\begin{equation}
\label{6.5}
J(\theta_s,\phi_s)=\frac{J_C+J_L}{J_L},
\end{equation}
where $J_L$ and $J_C$ are the contributions of the ladder and cyclic diagrams
\begin{equation}
\label{6.6}
J_L=\sum_{a=1}^A\sum_{j=1}^n\delta_j(\ks_F),
\end{equation}
\begin{equation}
\label{6.7}
J_C=\sum_{a=1}^A\sum_{j=2}^n\delta_j(\ks_F)\cos[{\bf q}\cdot{\bf R}_s^{(j)}].
\end{equation}
It should be recalled that the cyclic diagrams are formed starting from the double scattering. Here the summation over $a=1,2,\ldots A$ means the sum over all photons participating in the simulation. The summation over $j$ is performed over the orders of scattering. In the simulation we restrict ourselves by $n=10^5$ scattering orders. In addition if the contribution to the scattering of the order $\delta_j$ becomes very small, $\sim 10^{-8}$, then the higher orders for this photons are not taken into account.

\begin{figure}[h]
\begin{center}
\includegraphics[width=8cm]{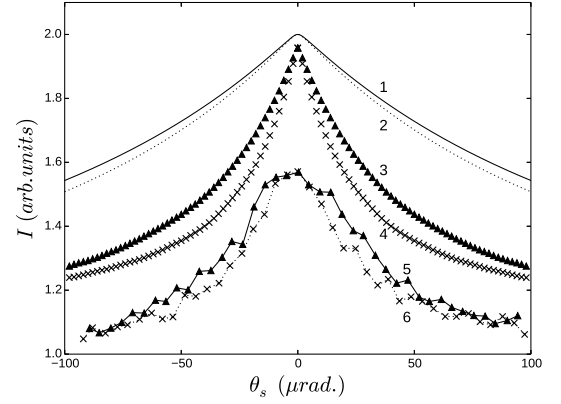}
\end{center}
\caption{Cross-sections of the coherent backscattering peaks.  Here curves (1) and (2) were obtained analytically~\cite{AksKuzRom2009}, curves (3) and (4) are the results of simulations,  curves (5) and (6) are the experimental data~\cite{Wiersma2004,Wiersma2005}. The  curves (1), (3) and (5) refer to the angle $\phi_s = 0$, the  curves (2), (4) and (6) refer to the angle $\phi_s = \pi/2$.
}
\label{fig11}
\end{figure}

The angular dependence   of the scattered radiation intensity was obtained from Eq.~(\ref{6.5}). Figure~\ref{fig11} shows the dependence of the intensity on the angle $\theta_s$ for two cross sections of the peak at $\phi_s={\pi}/{2}$ and $\phi_s=0$, curves 3 and 4. It is seen that there is a noticeable anisotropy of scattering. The calculations were performed for the liquid crystal 5CB studied in  works~\cite{Wiersma2005,Wiersma2004}. Parameters of this liquid crystal are $K_{11}=0.79 K_{33}$, $K_{22}=0.43 K_{33}$, $K_{33}=6.1\cdot10^{-7}$~dyne, $H=5000$~Oe, $\chi_a=1.38\cdot10^{-7}$, $\varepsilon_a=0.8$, $\varepsilon_\perp=2.2$, $\lambda=4.88\cdot10^{-5}$~cm, $T=301$~K. The experimental sample was a cylinder of 8~cm diameter and 4~cm  height. The extinction for such a liquid crystal is shown in Fig.~\ref{fig2}, curve 2 for the extraordinary beam and curve 4 for the ordinary beam. The figure shows that the mean free path of a photon is of order $l\sim 2\cdot 10^{-2}$~cm and it is significantly less than the sample size. Therefore in the simulation of the experiment  the approximation of the semi-infinite medium is quite justified.
Figure~\ref{fig11} also shows the results of analytical calculations, curves 1, 2 \cite{AksKuzRom2009}, and experimental data~\cite{Wiersma2005,Wiersma2004}, curves 5, 6.
In the selected geometry the contribution of the single scattering (e)$\rightarrow$(e) in the strictly backward direction  is absent.
In analytical calculations the (e)$\rightarrow$(e) was taking into account only.
The contributions of the ladder and the cyclic diagrams in this case are equal and the relative peak height should be equal to 2.
In numerical simulations the (o) and (e) beams are considered. Due to the presence of the (e)$\rightarrow$(o)  scattering the relative height of the peak becomes a little lower. Just such a result was obtained in our numerical simulation. In the experiment~\cite{Wiersma2005,Wiersma2004} this height was about 1.6.
Probably  this could be caused by the finite width of the instrumental function of the device. From Fig.~\ref{fig11} one can see that the results of the analytical calculations, lines 1, 2, predict peak width which differs from the numerical calculations. Probably the cause is  that in the summation of the diagram series  a number of assumptions have been made. Among them are the diffusion approximation~\cite{Stark1997} and  a simplified model for the pair correlation function.

\begin{figure}
\begin{center}
\includegraphics[width=8cm]{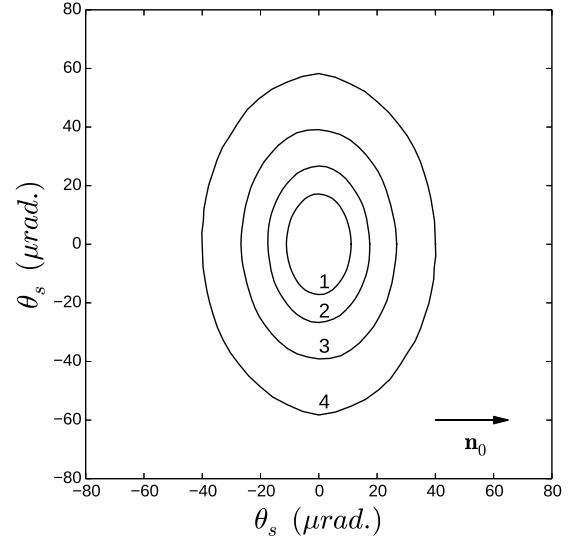}
\end{center}
\caption{Cross-sections of a coherent backscattering peak obtained by numerical
simulation in polar coordinates $(\theta_s,\phi_s)$. Cross-sections by planes perpendicular to the $z$ axis are calculated at
heights of $J = 1.7$ (1), $1.6$ (2), $1.5$ (3) and $1.4$ (4).
}
\label{fig12}
\end{figure}

The calculated anisotropy of the backscattering peak is shown in Fig.~\ref{fig12}. The cross-sections of the peak are shown at different heights: 1.7, 1.6, 1.5, 1.4. The calculated anisotropy of the peak is 1.46, in the experiment the anisotropy was $1.17\pm 0.04$.

\begin{figure}
\begin{center}
\includegraphics[width=8cm]{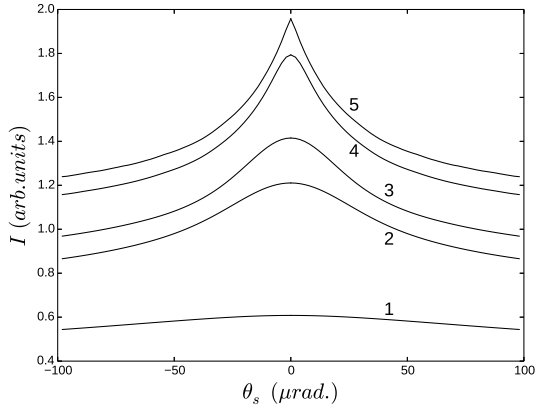}
\end{center}
\caption{The contribution of the first $n$  scattering orders to the
formation of the backscattering peak: (1) $n = 10$, (2) $n=50$, (3) $n=10^2$, (4) $n=10^3$, and (5) $n=10^5$.
All the curves are normalized by the sum of ladder diagrams that take
into account $n = 10^5$ scatterings.
}
\label{fig13}
\end{figure}

Performed numerical simulations allows to retrieve the details of the process which are practically impossible to obtain both experimentally and theoretically. In particular Fig.~\ref{fig13} shows the forming of the coherent backscattering peak when we take into account different number of scattering orders. All curves are normalized to the intensity determined by  summing of the ladder diagrams of all scattering orders, i.e. $n=10^5$. It is seen that the noticeable contribution made the lower orders of scattering. These contributions  are not described in the framework of the diffusion approximation as it was shown in the previous section. It means that the analytical calculations performed in the diffusion approximation can not fit the experimental curves with considerable accuracy.

\section*{Conclusion}

We have  investigated  the multiple scattering of  light in nematic liquid crystals with the aid of computer simulation method. The  main purpose was to carry out calculations without using any simplifying assumptions. It justifies our further quantitative comparison of the obtained results with the experimental data. The transition from radiative transfer in the form of contributions of separate scattering orders to the photon diffusion in the inhomogeneous media has been studied.  During simulation the non-monotonic dependence of the light diffusion coefficients on the external magnetic field has been obtained. Within the semi-analytical approach it was also possible to calculate the shape of the  coherent backscattering peak which appeared to be in a reasonable agreement with one obtained in the experiment.

The performed calculations illustrate the possibilities of computer simulation for study of the multiple scattering of waves in anisotropic systems with complex phase function. These results could be useful for description of the multiple scattering of elastic waves in the inhomogeneous media and, in particular, of the seismic waves. In this media both longitudinal and transverse elastic waves  propagate. Therefore accounting for the scattering in different types of the channels is principally important for an adequate description of the  multiple scattering in elastic media. The  performed calculations can be also useful for studying of the  light diffusion in biological objects where the true form of the phase function is particularly  important for the quantitative description of the experimental data. For such  systems the form of the phase function can be quite complicated  and  moreover the biological objects in some cases can have optical anisotropy the  account of which could be significant.

\begin{acknowledgments}
The authors acknowledge Saint Petersburg State University for a research grant \#11.38.45.2011 and the Russian Foundation for Basic Research for a grant \#12-02-01016-a. We thank Prof. P.N.~Vorontsov-Velyaminov for useful remarks.
\end{acknowledgments}

\end{document}